\documentclass[twocolumn,showpacs,preprintnumbers,amsmath,amssymb]{revtex4}

\usepackage{graphicx}
\usepackage{dcolumn}
\usepackage{bm}

\begin{document}

\preprint{}

\title{Multiple Superconducting Phases in New Heavy Fermion Superconductor PrOs$_4$Sb$_{12}$}

\author{K.~Izawa$^1$, Y.~Nakajima$^1$, J.~Goryo$^1$, Y.~Matsuda$^1$, S.~Osaki$^2$, H.~Sugawara$^2$, H.~Sato$^2$, P.~Thalmeier$^3$, and K.~Maki$^4$}

\affiliation{$^1$Institute for Solid State Physics, University of Tokyo, Kashiwanoha 5-1-5, Kashiwa, Chiba 277-8581, Japan}%
\affiliation{$^2$Department of Physics, Tokyo Metropolitan University, Hachioji, Tokyo 192-0397, Japan}%
\affiliation{$^3$Max-Planck-Institute for the Chemical Physics of Solid, N\"othnitzer Str.40, 01187 Dresden, Germany}
\affiliation{$^4$Department of Physics and Astronomy, University of Southern California, Los Angeles, CA 90089-0484}%


\begin{abstract}

The superconducting gap structure of recently discovered heavy fermion superconductor PrOs$_4$Sb$_{12}$ was investigated by using thermal transport measurements in magnetic field rotated relative to the crystal axes.  We demonstrate that a novel change in the symmetry of the superconducting gap function occurs deep inside the superconducting state, giving a clear indication of the presence of two distinct superconducting phases with twofold and fourfold symmetries.  We infer that the gap functions in both phases have a point node singularity, in contrast to the familiar line node singularity observed in almost all unconventional superconductors.   

\end{abstract}
\pacs{74.20.Rp, 74.25.Fy, 74.25.Jb, 74.70.Tx}
\maketitle

In almost all superconducting (SC) materials  known to date, once the energy gap in the spectrum of electrons opens at the SC transition, only its overall amplitude, and not the shape and symmetry around the Fermi surface, changes in the SC phase \cite{pok}.   The vast majority of superconductors have conventional $s$-wave pairing symmetry with an isotropic SC gap which is independent of directions over the entire Fermi surface.  Over last two decades, unconventional superconductivity with gap symmetries other than the $s$-wave has been found in several classes of materials, such as heavy fermion (HF) compounds\cite{sigrist}, high-$T_c$ cuprates\cite{tsuei}, ruthenate\cite{maeno}, and organic compounds\cite{kanoda}.  There, the strong Coulomb repulsion leads to a notable many-body effect and often gives rise to Cooper pair states with angular momentum greater than zero.  Unconventional superconductivity is characterized by anisotropic superconducting gap functions belonging to nontrivial representation of the crystal symmetry group which may have zeros (nodes) along certain directions.  The nodal structure is closely related to the pairing interaction of the electrons.  It is widely believed that the presence of nodes in almost all unconventional superconductors discovered until now are a signature of magnetically mediated interaction, instead of conventional electron-phonon mediated interaction.  A common feature in the HF superconductors discovered until now is that they all have {\it line} nodes in the gap functions, parallel or perpendicular to the basal planes. 

Recently a new HF superconductor PrOs$_4$Sb$_{12}$ with filled skutterudite structure has been discovered\cite{bauer}.  This material should be distinguished from the other unconventional superconductors, in that it has a non-magnetic ground state of the localized $f$-electrons in the crystalline electric field (most likely doublet $\Gamma_3$ state) \cite{bauer,vollmer,aoki4}.  The origin of HF behavior in this compound ($m*\sim50m_e$, $m_e$ is the free electron mass) was discussed in the light of the interaction of the electric quadrupole moments of Pr$^{3+}$, rather than local magnetic moments as in the other HF superconductors, with the conduction electrons.  Therefore the relation between the superconductivity and the orbital fluctuation of $f$-electron state  (i.e. quadrupole fluctuation) has aroused great interest; PrOs$_4$Sb$_{12}$ is a candidate for the first superconductor mediated neither by electron-phonon nor magnetic interactions.  Hence it is of the utmost importance to determined the symmetry of the SC gap.  Very recently heat capacity jump at $T_c$ in zero field was reported to be broad even in high quality samples\cite{vollmer}.  This behavior was attributed to the double superconducting transition, implying a possible novel feature of the superconducting state.    

Here we studied the SC gap structure of PrOs$_4$Sb$_{12}$ by the thermal transport measurements.  Cooper pairs do not carry entropy and therefore do not contribute to the thermal transport.  As a result,  thermal conductivity is a  powerful probe of the low energy excitations of unpaired quasiparticles (QPs).  Especially the measurements of the angular variation of the thermal transport in a magnetic field rotated relative to the crystalline axes  proved to be a uniquely powerful probe for determining the location of nodes \cite{yu,aubin,115,bedt,boro,ocana,vekhter,maki,vekhter3,won,tewordt,thalmeier}.   We report on what seems to be an exception to the well-known observation concerning the robustness of the symmetry of the SC gap function and the presence of line nodes, providing compelling evidence for the novelty of the SC state of PrOs$_4$Sb$_{12}$.

Our PrOs$_4$Sb$_{12}$ single crystals were grown by the Sb-flux method and had $T_c$=1.82~K with a very sharp resistive transition $\Delta T_c/T_c<0.01$.  Clear de Haas van Alphen oscillations were observed at 0.4~K,  confirming the high quality of the sample\cite{sugawara}.   In our samples, the heat capacity jump at $T_c$ in zero field showed a structure of some sort \cite{aoki2}, similar to Ref.\cite{vollmer}.  We measured the $c$-axis thermal conductivity $\kappa_{zz}$ (the heat current {\boldmath $q$}$\parallel c$) on the sample with a rectangular shape (0.40$\times$0.37$\times$2.20~mm$^3$) by the steady-state method.   To apply {\boldmath $H$} with high accuracy relative to the crystal axes, we used a system with two superconducting magnets generating {\boldmath $H$} in two mutually orthogonal directions and a $^3$He cryostat set on a mechanical rotating stage at the top of the Dewar.  By computer-controlling the two magnets and rotating stage, we were able to rotate  {\boldmath $H$} with a misalignment of less than 0.5$^{\circ}$ from each axis, which we confirmed by the simultaneous measurements of the resistivity. 

\begin{figure}[t]
\includegraphics[scale=0.5]{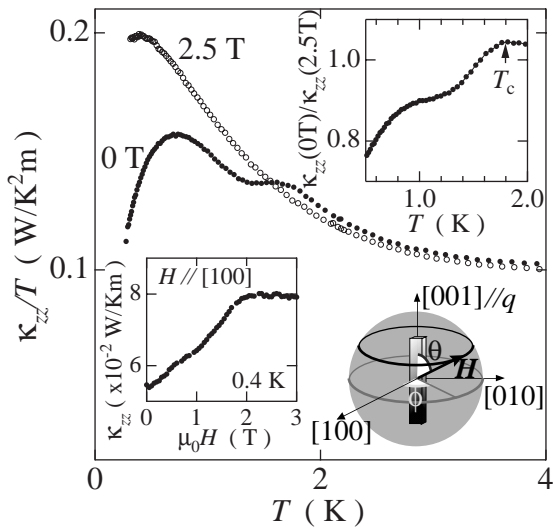}
\caption{Main panel: $T$-dependence of the $c$-axis thermal conductivity $\kappa_{zz}$ in zero field and above $H_{c2}$.  Insets; Upper: $T$-dependence of $\kappa_{zz}(H=0)/ \kappa_{zz}(H=2.5$~T) near $T_c$.   Lower left:  $H$-dependence of $\kappa_{zz}($ {\boldmath $H$}$\parallel $[100]) at 0.4~K.  Lower right:  schematic figure of the  {\boldmath $H$}=$H(\sin\theta\cos\phi, \sin\theta\sin\phi, \cos\theta)$ rotation.   While rotating $\phi$, $\theta$ was kept constant.}
\end{figure}

Figure 1 shows the $T$-dependence of $\kappa_{zz}$ in zero field and above $H_{c2}$ ($\simeq$ 2.2~T at $T$= 0~K).  In this temperature region, the electronic contribution to $\kappa_{zz}$ dominates  the phonon contribution.  In the upper inset of Fig.~1, $\kappa_{zz}$$(H=$ 0~T) normalized by $\kappa_{zz}$$(H=$ 2.5~T) is plotted.    Upon entering the SC state, a double shoulder structure is seen at $T \sim$1.6~K and 1~K, which may be related to the double SC transitions \cite{vollmer,aoki2}.  The lower left inset of Fig.~1 shows the $H$-dependence of $\kappa_{zz}$.   $\kappa_{zz}$ is nearly linear in $H$ over almost entire field range below $H_{c2}$.  This behavior is in a marked contrast to the exponentially slow increase of the thermal conductivity with field observed in $s$-wave superconductors at  $H\ll H_{c2}$.  The steep increase of $\kappa_{zz}$  in PrOs$_4$Sb$_{12}$ is a strong indication that the thermal transport  is governed by the delocalized QPs arising from the gap nodes\cite{boaknin}.

A strong additional evidence for the existence of the nodes is provided by the angular variation of $\kappa_{zz}$ in {\boldmath $H$} rotated within the $ab$-plane.  The theoretical understanding of the heat transport for superconductors with nodes has made significant advances during past few years\cite{hirsch}.   The most significant effect on the thermal transport for such superconductors comes from the Doppler shift of the QP energy spectrum ($\varepsilon(\mbox{\boldmath $p$})\rightarrow \varepsilon(\mbox{\boldmath $p$})-\mbox{\boldmath $v$}_s \cdot \mbox{\boldmath $p$}$) in the circulating supercurrent flow $\mbox{\boldmath $v$}_s$ \cite{volovik}.  This effect becomes important at such positions where the local energy gap becomes smaller than the Doppler shift term ($|\Delta| < |\mbox{\boldmath $v$}_s \cdot \mbox{\boldmath $p$}|$), which can be realized in the case of superconductors with nodes.  The maximal magnitude of the Doppler shift at a particular point strongly depends on the angle between the node direction and  {\boldmath $H$}.  For instance, when  {\boldmath $H$} is rotated within the basal plane in quasi two dimensional $d$-wave superconductors, the Doppler shift gives rise to the fourfold oscillation of the density of states (DOS).  In this case, the DOS shows the maximum (minimum) when  {\boldmath $H$} is applied to the antinodal (nodal) directions \cite{vekhter,won}.    This fourfold pattern is also present in the transport properties, and thus the thermal conductivity can be a powerful probe of the gap structure.  In fact, a clear fourfold modulation of the thermal conductivity, which reflects the angular position of nodes, has been observed in high-$T_c$ cuprate YBa$_2$Cu$_3$O$_7$\cite{yu,aubin,ocana}, HF CeCoIn$_5$ \cite{115}, organic $\kappa$-(ET)$_2$Cu(NCS)$_2$\cite{bedt},  and borocarbide YNi$_2$B$_2$C \cite{boro}, while it is absent in the fully gapped $s$-wave superconductors\cite{aubin}.   Here we apply this technique to PrOs$_4$Sb$_{12}$.

\begin{figure}[t]
\includegraphics[scale=0.43]{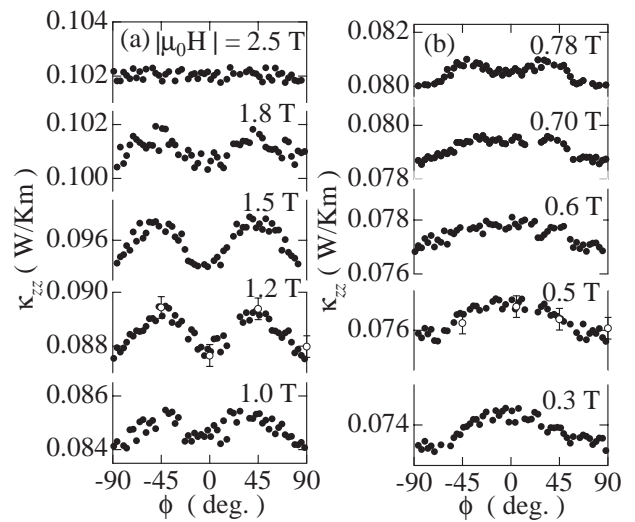}
\caption{(a)(b) Angular variation of $\kappa_{zz}$ in  {\boldmath $H$} rotating within the $ab$-plane ($\theta=90^{\circ}$ in the lower right inset of Fig.~1) at 0.52~K above and below $H_{c2}(\simeq2.0$~T).   }
\end{figure}

In the present experiments, $\kappa_{zz}$ was measured by rotating {\boldmath $H$}$=H(\sin\theta\cos\phi, \sin\theta\sin\phi, \cos\theta)$ conically around the $c$-axis as a function of $\phi$, keeping $\theta$ constant\cite{boro}.   Here $\theta$=({\boldmath $q$},{\boldmath $H$}) is the polar angle and $\phi$ is the azimuthal angle measured from the $a$-axis (see lower right inset of Fig.~1).  Figures 2 (a) and (b) display the angular variation of $\kappa_{zz}(${\boldmath $H$}$,\phi)$ in  {\boldmath $H$} rotating within the $ab$-plane ($\theta=90^{\circ}$) at $T$=0.52~K.  The measurements have been done in rotating  {\boldmath $H$} after field cooling at $\phi=-90^{\circ}$.  The open circles show $\kappa_{zz}(${\boldmath $H$}$,\phi)$ at $H$=1.2~T and 0.5~T which are obtained under the field cooling condition at each angle.  Values of $\kappa_{zz}(${\boldmath $H$}$,\phi)$ obtained by different procedures of field cooling are nearly identical, indicating that the field trapping effect related to the vortex pinning is negligibly small.  Above $H_{c2}$ ($\simeq$2.0~T at 0.5~K) $\kappa_{zz}(${\boldmath $H$}$,\phi)$ is essentially independent of $\phi$.  A clear fourfold variation is observed just below $H_{c2}$ down to $H \sim$ 0.8~T.  However further decrease of $H$ below 0.8~T causes a rapid decrease of the amplitude of the fourfold symmetry and eventually a discernible fourfold symmetry is not observed below 0.7~T.  At the same time, the twofold symmetry grows rapidly.    

\begin{figure}[t]
\includegraphics[scale=0.45]{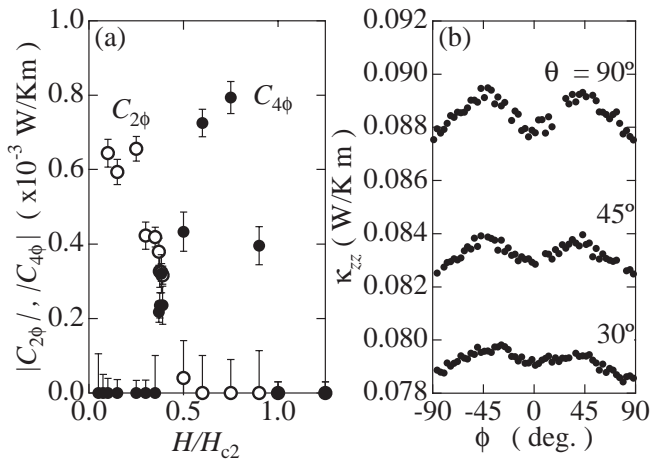}
\caption{(a)The amplitude of twofold (open circles) and fourfold (filled circles) symmetries plotted as a function of $H/H_{c2}$ at $T$=0.52~K.  For details, see the text. (b)Angular variation of $\kappa_{zz}(${\boldmath $H$}$,\phi)$ at $\theta=90^{\circ},45^{\circ}$ and $30^{\circ}$ measured by rotating  {\boldmath $H$} ( $|\mu_B${\boldmath $H$}$|$=1.2$~$T) conically around the $c$-axis as a function of $\phi$ (see the lower right inset of Fig.~1).}
\end{figure}

Figure~3(a) shows the $H$-dependence of the amplitude of twofold and fourfold symmetries.  The amplitude are obtained by decomposing $\kappa_{zz}(${\boldmath $H$}$,\phi)$ into three terms with different symmetries; $\kappa_{zz}(${\boldmath $H$}$,\phi)=\kappa_0+\kappa_{2\phi}+\kappa_{4\phi}$, where $\kappa_0$ is a $\phi$-independent term, $\kappa_{2\phi}=C_{2\phi}\cos2\phi$, and $\kappa_{4\phi}=C_{4\phi}\cos4\phi$ are terms with twofold and fourfold symmetry with respect to $\phi$-rotation.  We attempted to fit $\kappa_{zz}(${\boldmath $H$}$,\phi)$ using functions which possess the two and fourfold symmetries other than $\cos2\phi$ and $\cos4\phi$ and found that the $H$-dependence of both amplitudes are insensitive to the choice of the functions.  It is apparent that the transition from the fourfold to twofold symmetry in $\phi$-rotation occurs in a narrow field range at $H/H_{c2}\simeq$ 0.4, which is deep inside the SC state.  Both symmetries coexist in a narrow field range, possibly due to the inhomogeneity. 

It should be emphasized that the fourfold anisotropy of the Fermi velocity $v_F$, which is inherent to the cubic band structure of PrOs$_4$Sb$_{12}$, is quite unlikely to be an origin of the observed twofold and fourfold symmetries because of the following reasons.  First, the fact that $\kappa_{zz}(${\boldmath $H$}$,\phi)$ is $\phi$-independent above $H_{c2}$ strongly indicates that anisotropy of the Fermi surface does not produce a discernible anisotropy in the thermal transport.  Second, the observed twofold symmetry at lower field is lower than the symmetry of the Fermi surface.  Third, the amplitude of the anisotropy of $H_{c2}$ within the $ab$-plane, which is less than 1\%, is too small to explain the observed fourfold variation of $\kappa_{zz}(${\boldmath $H$}$,\phi)$  below $H_{c2}$.  Fourth, the calculations based on the Kubo formula indicate that the leading term in the thermal conductivity is related to the anisotropy of the gap function and anisotropy of $v_F$ will only enter as a secondary effect.  These consideration lead us to conclude that {\it the observed symmetries originate from the gap nodes}.  Therefore the change from fourfold to twofold symmetry provides a direct evidence for {\it the change of the gap symmetry}.  The fact that $\kappa_{zz}(${\boldmath $H$}$,\phi)$ possesses minima at $\phi=\pm90^{\circ}$ and 0$^{\circ}$ at high field phase and at $\phi=\pm90^{\circ}$ for low field phase immediately leads to the conclusion that the nodes are located along the [100]- and [010]-directions in the high field phase, while they are located only along the [010]-direction in the low field phase.

We discuss here why the lower twofold symmetry appears in the cubic lattice with higher symmetry.  The twofold symmetry indicates the absence of a double domain structure, suggesting a mechanism which fixes a single domain.  As shown in Fig.~2(b), the same twofold symmetry is observed in $\kappa_{zz}(${\boldmath $H$}$,\phi)$ measured under the field cooling condition at each angle.  This fact excludes the possibility in which the magnetic field fixes the domain.  Another mechanism, such as one dimensional dislocations produced in the process of the crystal growth, may be a possible origin for the single domain structure.  Further study is required to clarify its origin.

Having established the presence of nodes, the next question is their classification.  As demonstrated in Ref.\cite{boro,thalmeier}, the angular variation of $\kappa_{zz}$ can distinguish between the point and line nodes, by rotating  {\boldmath $H$}  conically around the $c$-axis with a tilted angle from the $ab$-plane.  In this case the amplitude of the fourfold and twofold symmetries decrease with decreasing $\theta$ for point nodes, while they essentially maintain the same magnitude for the line nodes.  This can be explained by noting that rotating  {\boldmath $H$} tilted from the $ab$-plane does not point to the nodes for the point node case, while rotating {\boldmath $H$} at any tilt angle always cross the nodes for the line node case.  Figure~3(b) displays the angular variation of $\kappa_{zz}$ at fixed $\theta$.  The amplitude at $\theta=45^{\circ}$ and 30$^{\circ}$ is smaller than that at $\theta=90^{\circ}$.  Similar results were obtained for the twofold symmetry.  These results are in favor of {\it point nodes}.  We note that these results are consistent with the recent NQR experiments which exclude the presence of the line nodes\cite{kotegawa}.

We next discuss the nodal structure inferred from the present results.  Unfortunately whether the point nodes are present along the [001]-direction remains unresolved.  This is because the measurements of $\kappa_{zz}$ as a function of the angle $\theta$ rotated across the $c$-axis is necessary to identify the nodes along the [001]-direction.   In such measurements, however, the dominant twofold oscillation in $\kappa_{zz}(\theta)$, arising from the difference in transport lifetime of QPs traveling parallel to and normal to the vortices \cite{vekhter,maki,won,tewordt}, obscures the part of the signal coming from the nodal structure along the [001]-direction.   We shall therefore be content here with a discussion of the nodal structure based on a group theoretical consideration.  It is unlikely that the SC gap function has four point nodes in the cubic $T_h$ crystal symmetry without taking into account the accidental mixing between the different basis functions in the SC wave function, regardless of spin singlet or triplet symmetry.  This naively suggests that  the gap function at high field phase has six point nodes, while that at low field phase has two point nodes.  More detailed study is necessary to clarify the nodal structure along [001]-direction.  

\begin{figure}[t]
\includegraphics[scale=0.55]{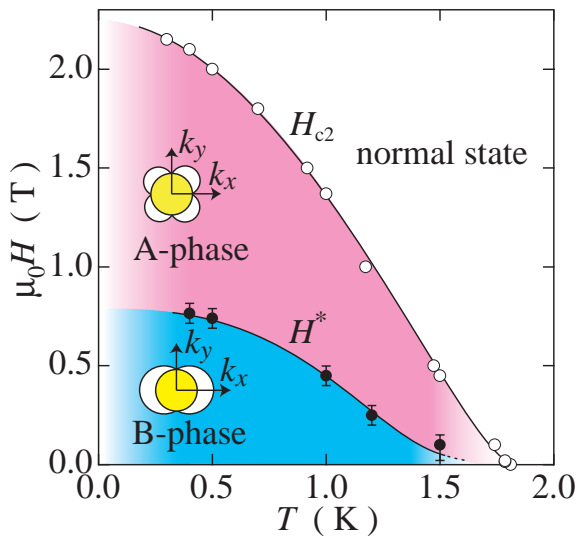}
\caption{The phase diagram of the superconducting gap symmetry determined by the present experiments.  The filled circles represent the magnetic field $H^*$ at which the transition from fourfold to twofold symmetry takes place.   The open circles represent $H_{c2}$.  The area of the gap function with fourfold symmetry is shown by pink ($A$-phase) and the area of the gap function with twofold symmetry is shown by blue ($B$-phase).   }
\end{figure}

The $H-T$ phase diagram of the SC symmetry determined by the present experiments is displayed in Fig.~4.    The filled circles represent the magnetic field $H^*$ at which the transition from fourfold to twofold symmetry takes place.    Near $T_c$ the amplitudes of both symmetries were strongly reduced but were sufficient to resolve $H^{*}$. The $H^{*}$-line which separates two SC phases (high field $A$-phase and low field $B$-phase) lies deep inside the SC state. This line seems to terminate at temperature slightly below $T_c$.   It is tempting to consider that the end point coincides with lower transition temperature of the double transitions discussed in Ref. \cite{vollmer}.  Its clarification motivates further investigations.  The only example of a superconductor with multiple phases of different gap symmetry so far has been UPt$_3$ \cite{sauls}.    However, the gap functions of UPt$_3$, in which the interaction leading to superconductivity is magnetic and  the antiferromagntic symmetry breaking field reduces the symmetry from hexagonal to orthorhombic, are known to be very complicated.  On the other hand, the SC gap functions of PrOs$_4$Sb$_{12}$  are expected to be much simpler owing to the cubic lattice symmetry.  

In summary,  we investigated the gap structure of heavy fermion superconductor PrOs$_4$Sb$_{12}$ using thermal transport measurements.  Two distinct SC phases with different symmetries, the phase transition between them, and the presence of point nodes-all these highly unusual nature highlight the unconventional superconductivity in PrOs$_4$Sb$_{12}$ and open up a new realm for the study of the superconductivity with multiple phases.  They also impose significant restrictions on possible theoretical models, including those with a novel mechanism of neither phonon nor magnetic origin of pairing.

We thank Y.~Aoki, H.~Harima, K.~Miyake, M.~Sigrist, K.~Ueda, and I.~Vekhter for helpful discussions.	

\begin{thebibliography}{50}
\bibitem{aoki4}The possibility of the non-magnetic singlet $\Gamma_1$ ground state is suggested in Ref.\cite{aoki3}.
\end{thebibliography}

%
\end{document}